\documentstyle[aps,psfig,preprint,floats]{revtex}
\tightenlines
\begin{document}
\bibliographystyle{normal}
\draft

\title{Phase-field Modeling of Eutectic Solidification:\\
From Oscillations to Invasion}

\author{Roger Folch$^*$ and Mathis Plapp}
\address{Laboratoire de Physique de la Mati\`ere Condens\'ee,
CNRS/\'Ecole Polytechnique, 91128 Palaiseau, France\\
$^*$ e-mail addresses: rf@pmc.polytechnique.fr, roger@gps.jussieu.fr, 
roger@ecm.ub.es
}
\maketitle
\begin{abstract}
We develop a phase-field model of eutectic growth
that uses three phase fields, admits strictly binary 
interfaces as stable solutions, and has a smooth 
free energy functional. We use this model to simulate
oscillatory limit cycles in two-dimensional lamellar
growth, and find a continuous evolution from 
low-amplitude oscillations to successive invasions 
of one solid phase by the other when the lamellar 
spacing is varied.   
\end{abstract}
\section{Introduction}
Solidification is both a 
fascinating example of pattern formation out of equilibrium 
and a phenomenon of practical importance in metallurgy.
The most common solidification microstructures found in 
industrial alloys are dendrites and eutectic composite 
structures (rods or lamellae). Eutectic growth occurs when 
two solid phases of different composition can solidify 
from the same melt. The interplay between the redistribution
of chemical components in front of the moving
phase boundary and the effects of capillarity along
the curved interfaces gives rise to a wealth of
different patterns and nonlinear phenomena such as bifurcations,
limit cycles, solitary waves, and even chaotic states \cite{ginibre}.

Phase-field modeling has become the method of choice for 
simulating solidification fronts. Its main advantage is 
that it avoids explicit tracking of the solidification front 
by introducing ``phase fields'' that vary continuously between
constant values corresponding to the bulk phases, thus replacing
the sharp fronts by diffuse interfaces with a finite thickness. 
The connection to the traditional free-boundary formulation 
of solidification is established by the technique of matched
asymptotic expansions around the equilibrium front in the
limit where the thickness of the diffuse interfaces is
much smaller than all other relevant length scales.
Recently, the computational efficiency of phase-field
models for single-phase solidification of pure
substances and binary alloys has been drastically 
enhanced by pushing this perturbation analysis to second 
order \cite{quantitative}. This, combined with a
random walk algorithm for an efficient simulation of the
diffusion equation, has made it possible to
simulate quantitatively the dendritic growth of a
pure substance in three dimensions for experimentally
relevant parameters \cite{rw}.

Therefore, it would be highly desirable to extend this 
second-order asymptotic analysis to phase-field models 
for multi-phase solidification; we will specifically
address the case of eutectic growth, where two solid
phases denoted by $\alpha$ and $\beta$ grow from the
liquid. The first phase-field models for eutectic
growth used the standard phase field to distinguish between
solid and liquid, and a concentration field \cite{alain,elder}
or a second phase field \cite{boettinger}
to distinguish between the two solids.
However, a solid-liquid interface then involves
variations of the two variables and the asymptotic 
analysis becomes extremely cumbersome.
The more recent multi-phase-field approach \cite{mpf} 
assigns one phase field to each phase and interprets 
these fields as local volume fractions. Across
an interface, in principle only the volume fractions of 
the two bulk phases limiting the interface need to vary, 
and one of them can be eliminated in terms of the other.
Hence, this method allows for equilibrium interface 
solutions depending on a single variable that are equivalent
to the usual ``binary'' interfaces of the standard
phase-field model. However, to achieve this 
the model has to be carefully designed, and
so far strictly binary interfaces have only been obtained
by using a singular free energy (double obstacle 
potential) \cite{2obstacle}, which might complicate 
the asymptotic analysis.

We present a phase-field model for eutectic growth 
that uses three phase fields, yields exactly binary 
interfaces at equilibrium, and has a smooth free 
energy functional. Furthermore, the free energy
is designed to keep the interfaces binary to first
order in the usual non-equilibrium asymptotic expansion,
which makes it a promising starting point for
a second-order analysis.

To illustrate the capabilities of the model in its
present form, we simulate lamellar eutectic growth of 
a simple model alloy with a symmetric phase diagram.
We show that small-amplitude oscillations of the 
lamellar width, ``giant oscillations'', and successive 
invasions of one solid phase by the other, each of
which is seen in experiments, all belong to 
the same branch of oscillatory limit cycles that
is parametrized by the lamellar spacing.

\section{Model}
We use three phase fields $p_i\in [0,1]$ representing the local
volume fractions of each phase ($\alpha$, $\beta$ or liquid),
and thus $\sum_i p_i =1$. Their dynamics is derived from 
a free energy functional $\cal F$,
\begin{equation}
\label{pi}
\frac{\partial p_i}{\partial t} = -\frac{1}{\tau} 
\frac{\delta {\cal F}}{\delta p_i}\enspace,
\end{equation}
where $\tau$ is a relaxation time and the variational
derivative has to take into account the constraint
$\sum_i p_i =1$.
Furthermore, the concentration field $c$
obeys the conserved dynamics
\begin{equation}
\label{c}
\partial c/\partial t = \vec\nabla\left[M(p_\alpha,p_\beta,p_l)
  \vec\nabla \mu \right] \enspace,
\end{equation}
where $\mu\equiv \delta{\cal F}/\delta c$ is the chemical
potential and $M$ is a phase-dependent mobility.

The free energy functional ${\cal F}=\int_V fdV$ is the
volume integral of a free energy density $f$ that is
conveniently decomposed into $f=f_{\rm grad}+f_{\rm TW}+f_c$,
where $f_{\rm TW}$ is a triple well potential with minima on 
each pure phase --- the equivalent of the double well potential 
of the standard phase-field model, $f_{\rm grad}$ contains
the gradients of the phase fields and ensures a finite
interface thickness and surface tension, and $f_c$ couples 
the phase fields and the concentration and drives the system 
out of equilibrium. 

\begin{figure}
\centerline{
 \psfig{file=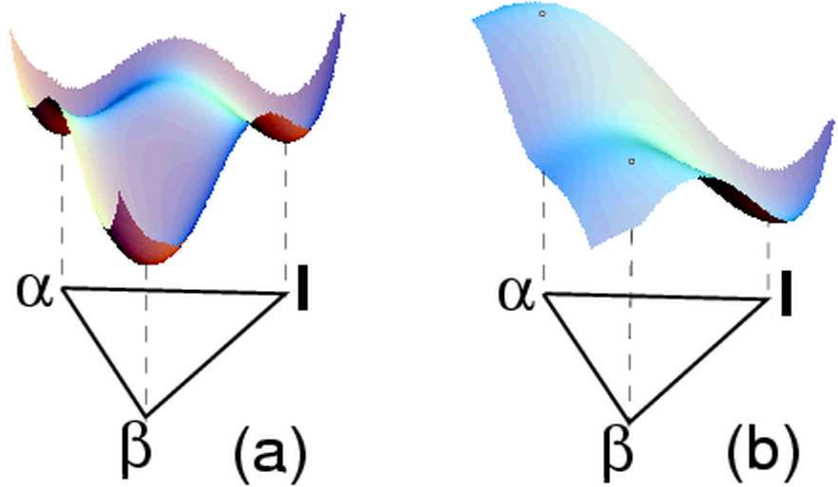,width=.8\textwidth}}
\medskip
\caption{Triple-well potential $f_{TW}$ (a) and elementary
tilt function $-g_l$ (b) drawn as ``landscapes'' over the
Gibbs simplex.}
\label{fig1}
\end{figure}

Due to the constraint $\sum_i p_i =1$, the free energy density 
and the interface solutions can be conveniently represented 
on a {\it Gibbs simplex}, that is an equilateral 
triangle where the distance to each side of the 
triangle from a given point represents the
value of one of the phase fields. Thus, each vertex 
corresponds to a pure phase, and each side to a 
purely binary interface. Our goal is to obtain
strictly binary interfaces, i.e., the equilibrium
solutions of (\ref{pi}) for an interface
connecting two phases should exactly run along
the edges of the simplex. One way to achieve
this is to choose
\begin{equation}
\label{gradient}
f_{\rm grad}=\frac{W^2}{2} \sum_i |\vec\nabla p_i|^2 \enspace,
\end{equation}
and to design the function $f_{TW}$ in such a way that 
the three minima are connected by ``saddles'' that run
along the edges and that have
vanishing derivatives in the direction
normal to the edges. Such functions can be
constructed using the geometry of the Gibbs simplex,
as will be detailed elsewhere. The simplest choice,
\begin{equation}
\label{triplewell}
f_{\rm TW}=\sum_i p_i^2 (1-p_i)^2 \enspace,
\end{equation}
is plotted in Fig.~\ref{fig1}a and reduces to the 
standard fourth-degree double well potential of 
the binary phase-field model on the edges.
With these choices, the free energy is completely
symmetric with respect to the interchange of any two
phases, and as a consequence all three surface 
tensions are equal. However, the general case of
unequal surface tensions can also be treated by
adding other terms in the free energy density
that modify the height of the saddle points.

Out of equilibrium, the triple well is ``tilted''
by the term $f_c$ in order to favor one phase over 
the others. Since we want to keep the interfaces 
binary, this term needs to be carefully constructed.
The elementary building blocks are tilt functions $g_i$
that satisfy $g_i=1$ for $p_i=1$, $g_i=0$ for $p_j=1$, 
$j\neq i$, and that have vanishing derivatives in the
direction normal to the edges of the simplex. One 
possible choice is
\begin{eqnarray}
\nonumber
g_i & = & 1 - \left[(1+p_j-p_i)^2 (2+p_i-p_j)^2
+ (1+p_k-p_i)^2 (2+p_i-p_k)^2\right. \\
\label{bascule}
&&\left. \mbox{} + (p_k-p_j)^2 (3-|p_k-p_j|)^2 \right]/12 \enspace.
\end{eqnarray}
In Fig.~\ref{fig1}b, we plot $-g_l$ that lowers the well for
the liquid with respect to the two others.
With the help of these functions,
\begin{equation}
\label{coupling}
f_c = \frac{1}{2}\left [c-\sum_i A_ig_i\right ]^2 + \sum_i B_ig_i
\end{equation}
interpolates between the concentration-dependent free energies
$f_i = (c-A_i)^2/2 + B_i$ of each phase $i$. Here, $c=(C-C_E)/\Delta C$
is a scaled concentration variable, where $\Delta C=C_\beta-C_\alpha$
is the width of the eutectic plateau, and $C_\alpha$, $C_\beta$, and
$C_E$ are the equilibrium compositions of the two solids and the
liquid at the eutectic temperature. An arbitrary phase
diagram can be constructed by performing the well-known double 
tangent construction, which yields
$c_{i}^{ij} = A_i + (B_j-B_i)/(A_j-A_i)$ for the equilibrium
concentration $c_i^{ij}$ of phase $i$ coexisting with phase $j$,
and by choosing appropriate $A_i(T)$, $B_i(T)$, where $T$ 
is the temperature. Note that by the same procedure, a
peritectic phase diagram can be constructed \cite{Lo01}.

Finally, we need to specify the mobility function in (\ref{c}).
Since $\partial^2 f_i/\partial c^2 \equiv 1$, the choice
$M(p_\alpha,p_\beta,p_l)=D p_l$ yields a constant diffusivity
$D$ in the liquid, and no diffusion in the solid (one-sided model).

\section{Simulations}
In thin-sample directional solidification of eutectic alloys
with global composition in the eutectic range, 
the basic pattern is a periodic array of alternating 
lamellae of each solid phase, growing perpendicular to the
large-scale solidification front
and parallel to the imposed temperature gradient.
This basic state exhibits various instabilities that 
have been carefully studied both experimentally in
the transparent organic alloy CBr$_4$-C$_2$Cl$_6$
\cite{ginibre} and by numerical simulations using
the boundary integral method \cite{KarSar}. For fixed 
experimental parameters, the stability of the
basic state is controlled by the width of one lamella 
pair, or {\it lamellar spacing}. For sample compositions 
close to the eutectic point, the first instability
that is encountered for increasing spacing is a
period-preserving oscillatory instability, that is,
the lamellae start to oscillate, with all the lamellae
of the same phase oscillating in phase, and the global
spacing left unchanged. Beyond the onset of this instability,
stable limit cycles are found, with an oscillation
amplitude that increases with spacing. In the boundary
integral simulations, this branch of solutions terminates
when the amplitude becomes too large because the thinner lamellae
pinch off. In contrast, in experiments ``giant oscillations''
with very large amplitudes can be obtained \cite{unpub}.
Furthermore, when the experiments are started from a
single solid phase growing into the liquid with a few 
widely spaced nuclei of the other solid phase on 
the solidification front, the nuclei
grow and spread along the front forming ``invasion
tongues'' \cite{invasion}. When two tongues growing
in opposite directions collide, they do not coalesce;
instead, a narrow channel of the other phase remains
and the whole process starts over with the role of
the two phases reversed. The interface dynamics is
hence still oscillatory, even if the patterns look
completely different.
In this regime, the boundary integral method becomes
inapplicable because it uses the quasistationary
approximation of the diffusion equation that is
valid only if the interface motion is slow compared
to the diffusive solute redistribution. 
Here, we investigate oscillatory limit cycles with our 
phase-field model, that does not have this restriction.

The sample is solidified with pulling
speed $V$ in a constant temperature gradient $G$
directed along the $z$ direction.
For simplicity, we use a symmetric phase diagram that
has constant concentration jumps across the interfaces
(parallel liquidus and solidus lines). This is
implemented by choosing $A_\alpha=-0.5$, $A_\beta=0.5$, 
$A_l=0$, $B_\alpha=B_\beta=0$ and $B_l=-(z-Vt)/l_T$,
where $l_T=(m\Delta C)/G$ is the thermal length, with
$m$ being the liquidus slope (equal for both phases).
The other relevant physical length scales are the
diffusion length $l_D=D/V$ and the capillary length
$d_0=\Gamma/(m\Delta C)$, where $\Gamma$ is the Gibbs-Thomson
constant. In terms of the model parameters,
$d_0=(2\sqrt{2}/3)W$. All simulations discussed here
were performed with $\tau=W=D=1$, $l_T/d_0=530$, $l_D/d_0=212$
and at exactly eutectic composition (that is, $c=0$
far ahead of the front).

The equations are integrated by a standard explicit
finite-difference scheme on a regular grid of spacing
$0.8$ $W$ for which we have checked that the discretization 
has converged. Since the diffusivity vanishes in the
solid, the diffusion equation only needs to be solved
in interfacial and liquid regions. Moreover, far ahead 
of the solidification front, it is solved using a
variant of the random walker algorithm of Ref.~\cite{rw}
that will be detailed elsewhere. Two types of
initial conditions are used. To obtain lamellar steady states,
simulations are started from two flat lamellae in contact with
the liquid. If the lamellar state is unstable, the
numerical noise generated by the random walkers
triggers the instability. Alternatively,
simulations are started from a flat single solid phase
at equilibrium with the liquid. Upon pulling, the
interface recoils towards colder temperatures as a 
diffusion layer of the rejected component is gradually 
formed. When the undercooling reaches a predetermined
value, nucleation is mimicked by placing a large enough 
semicircular nucleus of the other phase on top of the 
interface.

We restrict our attention here to period-preserving
oscillations without lateral drift. Therefore, it is
sufficient to simulate two adjacent half lamellae or half 
a nucleus of one solid phase on top of the other with 
reflecting (no-flux) boundary conditions on the sides 
of the simulation box that are parallel to the lamellae.
The lamellar spacing $\lambda$ is varied by changing the 
lateral size of the simulation box. In the figures below, 
for clarity we have reconstructed a whole lamella pair 
by reflecting the system with respect to one side 
of the box.

\begin{figure}
\centerline{
 \psfig{file=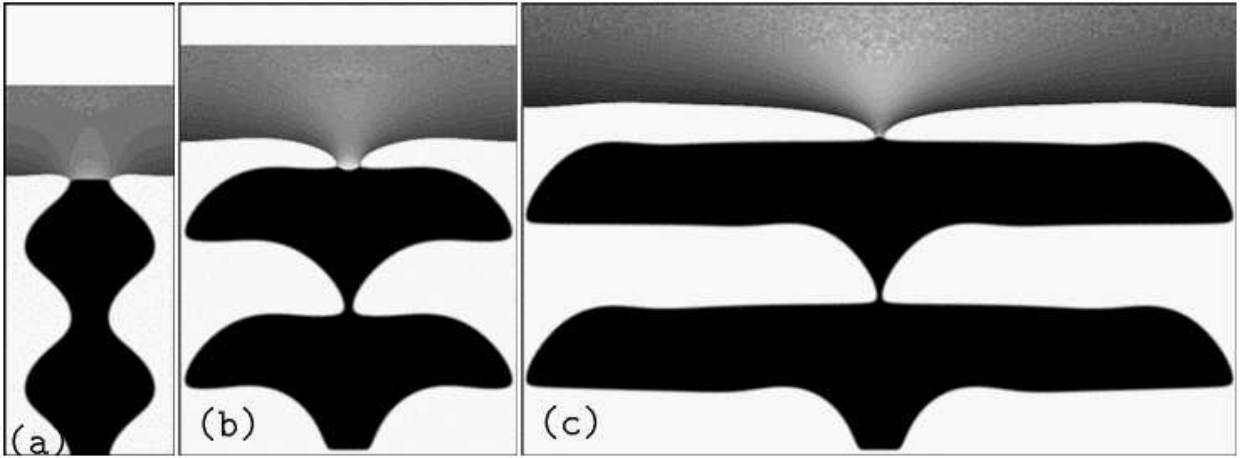,width=\textwidth}}
\medskip
\caption{Snapshot pictures of typical oscillatory structures.
Growth is upwards, the two solids are black and white, and
the greyscale in the liquid is proportional to the solute
concentration. (a) low-amplitude oscillations 
($\lambda/\lambda_{\rm min}=2$), (b) giant
oscillations ($\lambda/\lambda_{\rm min}=4$), 
(c) successive invasions ($\lambda/\lambda_{\rm min}=8.48$). 
The scale is the same on all three figures.
}
\label{fig2}
\end{figure}

We start by constructing the branch of stable lamellar
steady-state solutions and find the minimal undercooling 
spacing at $\lambda_{\rm min} \approx 85 d_0$, in good 
agreement with the prediction of the Jackson-Hunt theory \cite{JH}, 
$\lambda_{\rm min} \approx 79 d_0$. Next, we start from
a single nucleus and monitor the dynamics for increasing
spacings. We find a bifurcation toward oscillatory 
limit cycles for $\lambda$ close to $2\lambda_{\rm min}$.
As shown in Fig.~\ref{fig2}a, for $\lambda = 2\lambda_{\rm min}$
the oscillation already has an amplitude of about a quarter of
the lamellar spacing, but its shape is still close to a sinusoidal 
wave. At $\lambda = 4\lambda_{\rm min}$ (Fig.2~\ref{fig2}b), the 
oscillation is highly asymmetric, and has an amplitude
close to half the lamellar spacing, i.e., only a thin channel of
the other solid phase is left. Finally, 
at $8.48\lambda_{\rm min}$ (Fig.~\ref{fig2}c), 
three different regimes become apparent:
(i) initially, the nucleus grows slowly and spreads along 
the interface to form an invading finger with a 
well-defined shape, (ii) the finger speeds up with
an approximately constant acceleration, leaving a 
straight border with the other solid phase behind, and
(iii) when it approaches a finger growing in the
opposite direction (in our simulations, its mirror
image generated by the reflecting boundary conditions), 
it ``feels'' the diffusion field generated by
the other and slows down very rapidly, allowing for
the other phase to emerge through a narrow 
channel left between the two fingers and thus restart 
the process with the role of the two solids reversed.

Remarkably, our largest simulations reproduce many 
characteristic features of the ``successive invasions''
observed in the experiments \cite{invasion},
in particular the approximately constant acceleration 
of invasion tongues and the leftover of narrow 
channels between ``colliding'' fingers.
From our simulations, it
hence appears that this regime can be understood
as period-preserving oscillations with very large
amplitudes. Indeed, the morphology at $4\lambda_{\rm min}$ 
already presents stages (i) and (iii), but not (ii).
Apparently, to observe this intermediate, constant 
acceleration regime, it is sufficient to have 
enough ``free space'' left in front of the propagating
finger (i.e., large lamellar spacings). It is also
interesting to note that in peritectic alloys,
simulations of spreading in the same geometry
produce stages (i) and (ii), but then colliding 
fingers coalesce, and to obtain an oscillatory
dynamics explicit nucleation events have to be
introduced \cite{Lo01,Nestler00}.

The fact that we can switch from 
gentle oscillations to long invading fingers 
just by changing the lamellar spacing demonstrates 
that both belong to the same branch of oscillatory 
limit cycles. This branch of solutions is indeed an 
attractor, since simulations at fixed spacing and
started from very different initial conditions always
converge to the same cycle.

\section{Conclusion}
We have developed a phase-field model of eutectic
solidification that has a smooth free energy landscape
and yields exactly binary interfaces away from the
trijunction points. These properties make it
a promising starting point for a second-order asymptotic
analysis. Furthermore, we have used the model to
simulate oscillatory limit cycles in a model
alloy that has a symmetric phase diagram, and
we have found that low-amplitude oscillations,
``giant oscillations'', and successive invasions,
all observed in experiments, lie on a single
branch of solutions that is parametrized by the
lamellar spacing and that bifurcates from the
steady-state branch at about twice the minimum
undercooling spacing.

Here, we have focused on period-preserving oscillatory modes;
many other instability modes exist, for example
the tilt instability or period-doubling oscillations.
These instabilities can be studied using less 
restrictive boundary conditions. Preliminary simulations 
show that for the case studied here, namely an alloy 
with symmetric phase diagram at its eutectic 
composition, period-preserving
oscillations remain the only stable limit cycles.
In contrast, if we use off-eutectic sample
compositions or an asymmetric phase diagram, 
and hence break the complete symmetry between
the two solid phases, other modes become active.
Therefore, our model can be used to obtain a
more complete bifurcation diagram, which is
the subject of ongoing work.

We thank Silv{\`e}re Akamatsu and Gabriel Faivre
for many stimulating discussions and 
Jean-Fran{\c c}ois Gouyet for his help with Fig.~\ref{fig1}.
R. F. acknowledges financial support from project 
FMRX-CT96-0085 of the European Commission and from
Centre National d'{\'E}tudes Spatiales (France).

\end{document}